\documentstyle[12pt]{article} \begin{document} \thispagestyle{empty}
\begin{center} \LARGE \tt \bf {Cosmological magnetic field helicity and birefrigence from primordial torsion in Lorentz violation theories} \end{center}

\vspace{3.5cm}

\begin{center} {\large L.C. Garcia de Andrade \footnote{Departamento de F\'{\i}sica Te\'{o}rica - IF - UERJ - Rua S\~{a}o Francisco Xavier 524,
Rio de Janeiro, RJ, Maracan\~{a}}} \end{center}

\begin{abstract} Cosmological magnetic helicity has been thought to be a fundamental agent for magnetic field amplification in the universe. More recently Semikoz and Sokoloff [Phys Rev Lett 92 (2004): 131.301.] showed that the weakness of the seed fields did not necessarily imply the weakness of magnetic cosmological helicity. In this paper we present a derivation of dynamo equation based upon the flat torsion photon non-minimal coupling through Riemann-Cartan spacetime. From this derivation one computes the necessary conditions for a flat torsion field to generate a galactic dynamo seed, from the cosmological magnetic helicity. A peculiar feature of this dynamo equation is that the resistivity depends upon the Ricci scalar curvature. This feature is also present in turbulent dynamo models. Here the electrical effective conductivity is obtained by making use of flat torsion modes of a $R(\Gamma)F^{2}$ Lagrangean where R refers to the Ricci-Cartan spacetime. Power spectrum of the magnetic field is also computed. Lorentz violation appears naturally from birefrigence of photons semi-minimally coupled to torsion. Though Dobado and Maroto [Mod Phys Lett A 12: 3003 (1997)] have previously investigated the role of primordial torsion in the anisotropy of light propagation they made it using the fermionic sector of the QED Lagrangean while we obtained similar results using the photonic sector. They also used the pseudo-trace of torsion while we here work out with the torsion trace itself.

\end{abstract} Key-words: Early Universe, Lorentz violation, torsion theories,
cosmology, astro-particle physics. \newpage

\section{Introduction} Earlier Ahonen  \cite{1} computed the electrical conductivity in the early universe from the Riemannian version of the Boltzmann equation, taking into account cosmic scales. In this paper one computes the effective conductivity from the electrical conductivity in flat space perturbed by flat torsion modes obtained from the non-minimal coupling of the e.m field and Ricci scalar. Turner and Widrow \cite{2} have investigated this lagrangean in the context of galactic dynamo seeds. In Riemannian spacetime of general relativity they argued that this Lagrangean yields weak primordial fields which are not strong enough to seed galactic dynamo.
In this paper one derives the dynamo equation from this photonic sector of QED \cite{3} and obtain electrical conductivity without using the Boltzmann equation in Riemann-Cartan spacetime. Dynamo equation in anisotropic Bianchi type I model has been investigated by . Power spectra of magnetic fields and magnetic helicity, are computed.  The paper is organised as follows: In section II we
compute dynamo equation and the curvature dependent electrical conductivity and the power spectra of the magnetic field. The helicity of magnetic field is also computed and consequences for galactic dynamo seeds discussed. In this same section the the influence of torsion fields on the birefrigence \cite{4} in the propagation of e.m fields is also presented. Section III contains
conclusions and discussions.
\section{Galactic dynamo seeds and birefrigence from flat semi-minimal torsion-photon coupling
of $RF^{2}$ Lagrangean}
Though torsion effects are highly suppressed in comparison with curvature
ones of Einstein gravity sector, we do not consider here Minkowski space since as can be easily shown here from the field equations that torsion vanishes in Minkowski space. Turner and Widrow Lagrangean \cite{2} is
\begin{equation}
S=\frac{1}{m^2}\int{d^{4}x(-g)^{\frac{1}{2}}(-\frac{1}{4}[1-4R(\Gamma)]F^{2}])}
\label{1}
\end{equation}
where ${\Gamma}$ is the Riemann-Cartan connection. Euler-Lagrange equations yields the following field equations:
The Maxwell equations
\begin{equation}
{\partial}_{\mu}F^{\mu\nu}=F^{\mu\nu}\frac{{\partial}_{\mu}R}{(1-R)}
\label{2}
\end{equation}
and the Bianchi identities
\begin{equation}
{\partial}_{[\mu}F_{\alpha\nu]}=0
\label{3}
\end{equation}
and the torsion trace $T_{\nu}={T^{\mu}}_{\nu\mu}$. Here we shall assume that only the time component of the torsion trace survives. This time component is represented simply by T. Thus the torsion trace equation is
\begin{equation}
\Box{T}-T={\partial}_{\eta}(E^{2}-B^{2})
\label{4}
\end{equation}
where ${\eta}$ is the conformal coordinate of Minkowski space given by the line element
\begin{equation}
ds^{2}=a^{2}(\eta)(d{\eta}^{2}-d\textbf{x}^{2})\label{5}
\end{equation}
By assuming the following ansatz
\begin{equation}
\textbf{B}_{k}=\textbf{B}_{0}e^{{\omega}_{B}{\eta}}\label{6}
\end{equation}
\begin{equation}
\textbf{E}_{k}=\textbf{E}_{0}e^{{\omega}_{E}{\eta}}\label{7}
\end{equation}
and
\begin{equation}
\textbf{T}_{k}=\textbf{T}_{0}e^{{\omega}_{T}{\eta}}\label{8}
\end{equation}
Fourier transforming the torsion equation and substitution of the last three equations yields
\begin{equation}
T_{k}=\frac{2{\omega}({E_{0}}^{2}-{B_{0}}^{2})}{(1 + k^{2}+4{{\omega}}^{2})}e^{2{\omega}\eta}
\label{9}
\end{equation}
where ${\omega}={\omega}_{B}={\omega}_{E}=\frac{1}{2}{\omega}_{T}$ is a kind of degeneracy in the frequency. Let us now consider the Maxwell equations in the form
\begin{equation}
{\partial}_{\eta}\textbf{B}+{\nabla}{\times}\textbf{E}=0
\label{10}
\end{equation}
\begin{equation}
{\partial}_{\eta}\textbf{E}+{\nabla}{\times}\textbf{B}=\frac{\dot{R}}{(1-R)}\textbf{E}
\label{11}
\end{equation}
\begin{equation}
{\nabla}.\textbf{B}=0
\label{12}
\end{equation}
\begin{equation}
{\nabla}.\textbf{E}=0
\label{13}
\end{equation}
Taking the curl of equation (\ref{10}) and substitution of equation (\ref{9}) yields
\begin{equation}
-{{\partial}_{\eta}}^{2}\textbf{B}+{\nabla}^{2}\textbf{B}=-\frac{\dot{R}}{(1-R)}{\partial}_{\eta}\textbf{B}
\label{14}
\end{equation}
Computing the Fourier spectrum it yields
\begin{equation}
{{\partial}_{\eta}}^{2}\textbf{B}_{k}+{k}^{2}\textbf{B}_{k}=\frac{\dot{R}}{(1-R)}{\partial}_{\eta}\textbf{B}_{k}
\label{15}
\end{equation}
By considering the plasma effects one has for the electrical conductivity
\begin{equation}
{\sigma}=\frac{\dot{R}}{(1-R)}
\label{16}
\end{equation}
Note that in the early universe where torsion or curvature effects are strong the electrical conductivity is negative and the conductivity is negative and the effective conductivity decreases which means more electric resistivity in the cosmic plasma due to torsion effects. By making use of the above ansatz for the magnetic field and substituting into expression (\ref{10}) one obtains
\begin{equation}
{\omega}^{2}+k^{2}-\frac{\dot{R}}{(1-R)}{\omega}=0
\label{17}
\end{equation}
Since the phase velocity is given by $v_{ph}=\frac{{\omega}}{k}$ the dispersion relation (\ref{16}) shows the well known result that the photons propagated with two distinct polarisation states. Dobado and Maroto \cite{4} have proven this in the torsion context, and birefrigence is also present here. Nevertheless they worked out in the fermionic sector of QED and our approach is in the photonic sector of the QED with torsion. They also consider a tiny torsion to explain birefrigence. This fact allow us a simple interpretation of Lorentz violation associated with the torsion theory here as done previously by Kostelecky et al \cite{5}. The dispersion relation yields
\begin{equation}
{\omega}_{\pm}= \frac{{\partial}_{\eta}R}{[1-R]}{\pm}ik
\label{18}
\end{equation}
Note that this result reduces to the vacuum Maxwell equation one when the Ricci-Cartan scalar vanishes. The helicity expression
\begin{equation}
{\cal{H}}_{k}=\frac{k^{2}}{2{\pi}^{2}}(|{B^{+}}_{k}|^{2}-|{B^{-}}_{k}|^{2})
\label{19}
\end{equation}
Recently Semikoz and Sokoloff \cite{6} have discussed and contest the paradigm that the weak field that seeds the magnetic galactic dynamo is weak which implies a weak helicity. Let us now consider the computation of the last expression helicity in terms of torsion to check if the flat torsion modes induce a weak or strong helicity.
Before computing the helicity let us consider the comoving dissipation length
\begin{equation}
{{\epsilon}^{2}}_{diss}=\int{\frac{d{\eta}}{\sigma}}
\label{20}
\end{equation}
For modes well within the horizon, $k|\eta|>>>1$ then the expression
\begin{equation}
{{\partial}^{2}}_{\eta}\textbf{B}_{k}+k^{2}\textbf{B}_{k}={\sigma}{\partial}_{\eta}\textbf{B}_{k}
\label{21}
\end{equation}
reduces to
\begin{equation}
k^{2}\textbf{B}_{k}\approx{{\sigma}{\partial}_{\eta}\textbf{B}_{k}}
\label{22}
\end{equation}
A simple solution of this equation is given by
\begin{equation}
{B^{\pm}}_{k}\approx{{{B^{\pm}}^{0}}_{k}exp[\int{d{\eta}\frac{k^{2}}{\sigma}}]}
\label{23}
\end{equation}
From the definition of dissipation length above yields
\begin{equation}
{B^{\pm}}_{k}\approx{{{B^{\pm}}^{0}}_{k}exp[-{\epsilon}^{2}k^{2}]}
\label{24}
\end{equation}
Since the dissipation length ${{\epsilon}_{diss}}^{2}\sim{\frac{1}{{\sigma}H}}$, where H is the Hubble constant, where $\sigma\sim{\frac{1}{H}}$. The magnetic helicity is then given by
\begin{equation}
{\cal{H}}_{k}\sim{exp[-{\epsilon}^{2}k^{2}]}
\label{25}
\end{equation}
Therefore unless $R>>> k\eta$ or the torsion is extremely strong, the magnetic field helicity is washed out and the
magnetic field presents a fast decay and the cosmological magnetic field is not strong enough to seed galactic dynamos. Note that the dynamo equation can be obtained from the expression
\begin{equation}
{\nabla}^{2}\textbf{B}_{k}-{{\sigma}{\partial}_{\eta}\textbf{B}_{k}}=0
\label{26}
\end{equation}
The power spectra of the magnetic field
\begin{equation}
{\cal{P}}_{k}=\frac{k^{3}}{2{\pi}^{2}}(|{B^{+}}_{k}|^{2}+|{B^{-}}_{k}|^{2})
\label{27}
\end{equation}
which shows that power spectra decay fast unless the torsion is extremely strong such as in black holes or in the very early universe.

\section{Discussion and conclusions}
In this paper we show that a semi-minimal torsion coupling in Riemann-Cartan spacetime can be used to generalize previous papers on QED cosmology where a Minkowski space with torsion has been used.  As Lagrangeans can be used to determine the torsion which can be used to seed galactic dynamos. The motivation from this
study came from some work by Campanelli et al \cite{7} where they investigate
similar subjects in the general relativistic backgrounds of Riemannian geometry. We conclude that in order that a magnetic field may be generated from the primordial torsion fields it has to be extremely strong, while a tiny torsion field would be enough to generate anisotropy in the propagation of the e.m fields as shown by A Dobado and Maroto \cite{4} many years ago in the context of fermionic sector of Lorentz violated QED. Generalisations of the work presented here for the general curved and torsioned spacetime, may appear elsewhere.
\section{Acknowledgements}
 We would like to express our gratitude to A Maroto and D Sokoloff for
helpful discussions on the subject of this paper. Special thanks go to an anonymous referee for his extremely kind suggestions which allowed us to making considerable and important improvements on the first draft of this letter. One of us (GdA) would like to thank financial supports from CNPq. and University of State of Rio de Janeiro (UERJ). 
\end{document}